\documentclass[usenatbib]{mn2e}
\usepackage{epsfig}
\def\apj{ApJ}
\newcommand{\beq}{\begin{equation}}
\newcommand{\eeq}{\end{equation}}

\def\rs{R_{\rm smooth}}
\def\msun{{\rm M}_\odot}
\title{Fast Identification of Bound Structures in Large N-Body Simulations}
\author[Jochen Weller, Jeremiah P Ostriker, Paul Bode and Laurie Shaw]
{J.~Weller$^1$\thanks{Email: J.Weller@ast.cam.ac.uk}, J.~P.~Ostriker$^{1,2}$, P.~Bode$^2$ and L.~Shaw$^1$\\ $^1$ Institute of
Astronomy, University of Cambridge, Madingley Road, Cambridge CB3
0HA.\\ $^2$ Princeton University Observatory, Princeton NJ 08544-1001.}
\date{Accepted ???, Received ???; in original form \today}
\pagerange{\pageref{firstpage}--\pageref{lastpage}}
\pubyear{2003}

\begin{document}
\maketitle
\label{firstpage}
\begin{abstract}
We present an algorithm which is designed to allow the efficient identification and
preliminary dynamical analysis of thousands of structures and
substructures in large N-body simulations. 
First we utilise a refined density gradient system (based on {\rm Denmax})
to identify the structures, and then apply an
iterative approximate method to identify unbound particles, 
allowing fast calculation of bound substructures. 
After producing a catalog of separate energetically bound
substructures we check to see which of these are energetically bound
to adjacent substructures.  For such bound complex subhalos, we combine
components and check if additional free particles are also bound to the
union, repeating the process iteratively until no further changes are
found.  Thus our subhalos can contain more than one density maximum,
but the scheme is stable:  starting with a small smoothing length
initially produces small structures which must be combined later, and
starting with a large smoothing length produces large structures
within which sub-substructure is found. We apply this algorithm to
three simulations. Two which are using the TPM algorithm by
\citet{Bode:00a} and one on a simulated halo by \citet{Diemand:04b}.
For all these halos we find about 5-8\% of the mass in substructures.

\end{abstract}


\section{Introduction}
Until recently,
observational extra-galactic astronomy has been
based primarily on the study of galaxies and clusters of
galaxies. The theoretical constructs 
in the standard $\Lambda$CDM paradigm for structure formation
which are most closely associated with these phenomena 
are ``halos'' of dark matter and the ``subhalos''
within them.
In this bottom up picture, all self gravitating virialised
objects are comprised of accumulated smaller objects, and these latter,
hierarchically, of still smaller ones {\em ad infinitum}, 
assembled through ``merger trees''. Thus a close examination of any
representative object should show the undigested remnant cores of
previously ingested objects, tidal streamers of debris shredded from
the outer parts of these same subhalos, and the relatively smooth
background material which contains the somewhat phase mixed accumulation of
all the digested tidal effluvia. A closer and closer analysis in phase
space would allow identification of components added at earlier and
earlier times.

Thus ``identification of substructure'', even if perfect tools were
available, requires some intellectual precision in the dynamical
definitions of what is meant by ``subhalos''. Until recently the lack
of sufficiently accurate computations made this issue moot, but now
investigators have begun this analysis, using a variety of defined
terms. We will provide our own definitions later in this section.

Historically, it was impossible to produce galaxy-size halos in
dense clusters with dark matter simulations
\citep{White:76a,vanKampen:95a,Summers:95a,Moore:96a}. This was mainly
due to the limited mass and force resolution of 
the simulations used and was commonly known as the {\em over-merging
problem}. The major causes of this problem
were premature tidal disruption due
to inadequate force resolution and 
two-particle evaporation for halos with a small number of particles
\citep{Klypin:99a}. However the combination of an increase in
computing power and the invention of more efficient algorithms has led
to promising developments over the recent years which have overcome
the numerical problems \citep{Ghigna:98a,Klypin:99a,Moore:99a,Okamoto:99,Ghigna:00a,Bode:00a,
Springel:01a,DeLucia:03,Kravtsov:03}.
Besides the numerical insufficiencies which can 
destroy substructures, there are also physical reasons for the
destruction of structure, which are tightly connected to the
numerical problems. First, there is dynamical friction, which drives the
subhalo to the halo centre where it can be disrupted and merge with
the central  
object. Second, there is tidal stripping when the tidal force from
the halo on the subhalo is larger than the gravitational force 
holding the subhalo together. 
Furthermore, there may be shock heating which occurs during
the close passage of two subhalos, and more dominantly on passing of a
subhalo near the halo centre; this effect is believed to be less
prominent than the first two \citep{Moore:96a,Klypin:99a, Gnedin:99a}.

\cite{Klypin:99a} estimated that a force softening of $\epsilon=3\,
h^{-1}\; {\rm kpc}$ and a mass resolution below
$m_p=10^9\,h^{-1}\;\msun$ would be sufficient to identify a substructure
of mass $10^{11}\,h^{-1}\,\msun$ with at least 30 particles at a
distance $70h^{-1}{\rm kpc}$ from the centre of a
$10^{14}h^{-1} \msun$ cluster. Needless to say, higher resolution would 
be even better. The usual approach to obtain such resolution
is to take a cluster from a cosmological
N-body simulation and re-simulate it at higher resolution with
inclusion of the long distance (tidal) gravitational fields. However if one
wants to address the problem of substructure in a statistical and 
cosmological context, then one needs fairly large simulation boxes. 
Thus one
cannot currently use, with existing computing power, much higher 
resolution than given above. 

Our goal is to design algorithms that can be used to analyze
structure/substructure in very large simulations such as ``light cone
radians'' of the Virgo group \citep{Evrard:02b} rather than
individual very high resolution simulations of clusters.
Besides the noted numerical difficulties, the identification of
structures and substructures in large N-body simulations is a long
standing problem of principle. This has been addressed in the 
past by many different methods, mainly geometrical rather than physical
\citep{Huchra:82a,Davis:85a,Bertschinger:91a,Gelb:94a,Warren:92a,Lacey:94a,
Stadel:97a,Weinberg:97a,Eisenstein:98a,Klypin:99a,
Springel:01a}. Many methods exploit to some extent the
friends-of-friends (FOF) \citep{Huchra:82a,Davis:85a,Lacey:94a} or the
Denmax \citep{Bertschinger:91a,Gelb:94a,Eisenstein:98a} algorithm
(described in Section \ref{sec:method}), which are also at the
centre of our method.
What these
methods have in common is that they are essentially geometrical and do not
use the entire phase space information, and hence need post processing
to test for bound structures. In this paper we discuss a fast approximate 
method to remove unbound particles from halos.

Algorithms which have been used for finding bound structure
include SKID \citep{Stadel:97a} and hierarchical
adaptations of it, BDM \citep{Klypin:99a}, and
SUBFIND \citep{Springel:01a}. 
Quite recently other methods have been introduced by
\citet{Kim:04}, \citet{Neyrinck:04}, and \citet{Knebe:04}. 
SKID essentially uses the Denmax 
algorithm to identify structures, and then calculates bound structures
by iteratively removing the unbound particle with the largest total
energy until all particles are bound. The hierarchical scheme
\citep{Ghigna:00a} uses SKID at three different smoothing
lengths. The BDM (bound density maximum) scheme places spheres of
a certain scale $r_{\rm sp}$ in 
the simulation box, and then displaces the spheres to the centre of
mass of the particles inside the sphere. This process is iterated and
eventually all maxima within a sphere of size $r_{\rm sp}$ are
found. The unbinding is then done by calculating the escape velocity
of the halo from the maximal circular velocity; all
particles with velocities larger than the escape velocity are
removed. For the calculation of the escape velocity a
Navarro-Frenk-White (NFW) density profile is assumed
\citep{Navarro:95a}. Recently BDM has been used to identify a vast
number of halos in a
large detailed simulation \citep{Kravtsov:03}. 
The SUBFIND algorithm uses the FOF
algorithm to find cluster-sized halos, and then looks for saddle points
in the density field to identify subhalos. Again, the particles of a
subhalo are then examined to determine if they are bound. Recently
11 re-simulated clusters have been analysed in great detail with
this method \citep{DeLucia:03}.

Most of the work quoted above used the re-sampling technique and
consequently  only analysed a small number of ``typical halos'' to
high accuracy.  Here we take a complementary approach, by using simulations
of volumes containing many target halos.
While sacrificing resolution (as compared to the re-sampling
technique) we gain in sample size by a large factor, with thousands of
halos in our largest runs. In this paper we will study
two quite different simulations.
One contains $256^3$ particles in a volume 
$20 h^{-1}\,{\rm Mpc}$ on a side; the halos from this run have masses
typical of large galaxies.  This run is discussed in more detail
in \citet{Bode:01}; it was evolved with a P$^3$M code, and halted
at redshift $z$=1.  The second simulation is of $1024^3$ particles
with box size $320 h^{-1}\,{\rm Mpc}$, containing many galaxy
cluster sized halos.  This was evolved to $z$=0.05 using the
Tree-Particle-Mesh (TPM) algorithm \citep{Bode:03a}. The simulation
parameters can be found in Table \ref{tab:simulations}. One
  difference between the two codes used is that P$^3$M uses Plummer, and TPM uses spline, softening. 

\begin{table*}
\begin{tabular}{l|ccccccccccc}
Model & z &$\Omega_b$ & $\Omega_c$ & $\Omega_\Lambda$ &
$H_0\,\left[\frac{\rm km/sec}{\rm Mpc}\right]$ &
$\sigma_8$ & $n$ & $N$ & $L$ [Mpc/h] & $m_p \; [M_\odot/h]$ & $\epsilon$
[kpc/h]\\
\hline 
$\Lambda$CDM  &  1.0 & 0.04 &   0.26 &   0.70 &   67  &    0.900  &  1.0  &   $256^3$
&  20 & $3.97 \times 10^7$ & 1.2 \\
$\Lambda$CDM & 0.05 &  0.04 &   0.26 &   0.70  &  70  &    0.975 &   1.0 &
$1024^3$ &  320 & $2.54 \times 10^9$ & 3.2 \\
\hline
VIRGO & & & 1.0 & 0.0 & 50 & 0.7 & & 1314161 & & $8.6 \times 10^8$ & 5,10 \\  
\hline
\end{tabular}
\caption{In the top two rows the parameters for the two simulations analysed in this paper, cold dark matter universes with a cosmological constant. In the last row the parameters for the re-simulated cluster by \citep{Ghigna:98a}.}
\label{tab:simulations} 
\end{table*}

We will define a subhalo at any level of the hierarchy in the
following fashion.
In the centre of mass frame defined by the object in question, we take
all particles as members which are gravitationally bound
($E<0$). Thus, if a small smoothing length has been used to identify
sub-clumps, we check if groups of these are bound to one another and
if additional ``free'' particles are bound to the
assemblage. Conversely, if a larger smoothing length has been used to
identify objects we subsequently analyse these with greater refinement
to ascertain subcomponents which in their own frames are self-gravitating. 
Thus we produce a catalog which provides labels for a
hierarchy of bound objects, where the catalog is, to a large extent,
independent of the geometrical tools used to parse the entire
object. We then make an independent catalog of the hierarchy,
where at each level we require all components to be gravitationally bound to the object to which they are attached.

The purpose of this paper is to clearly define the method and
attempt to
carefully specify the algorithms that define and identify substructure
and to explain how seemingly minor variations in procedure can produce
large changes in the final result. 

\section{The method}\label{sec:method}
Before entering into the details of the method, we present a schematic overview of
the substructure finding algorithm which we will employ. We first
apply the FOF method,
which groups together large structures in a
speedy way, on the entire simulation volume.  At the core of our
approach is the geometrically based Denmax routine by 
Bertschinger \& Gelb (1991), which moves particles up density gradients and
identifies groups as all particles reaching the same density maximum. 
We run Denmax with high resolution on each FOF halo.
We then  
build a family tree and identify, with an iterative approximation
scheme, energetically bound
particles within the structures. 

In this way we create, hierarchically, (a) gravitationally bound
objects (``mothers''), (b) those substructures which lie within a given
bound object (``daughters'')
and are themselves gravitationally bound, and (c) further sub-levels.

\subsection{Creating the family trees}
The first step to identify large groups in the simulation is the
application of the FOF routine.
We choose as a linking length $R_{\rm link} = 0.2{\bar n}^{-1/3}$,
where ${\bar n}^{-1/3}$ is the mean inter-particle separation.
This ensures that we find clusters, and also 
trace them out to the virial radius. Furthermore, this choice will select
groups of particles with over-densities
close to the value predicted by the spherical
collapse model. With this linking length, and a minimum number of ten
particles required to be identified as a group, FOF finds a large
number of low mass halos and a decreasing number of more massive objects.

We estimate the
density at each point of the simulation by measuring the weighted volume over
the 16 nearest neighbours of each particle in the
simulation (using the SMOOTH code; see
http://www-hpcc.astro.washington.edu/tools). 
This enables us to estimate the position of the density peak of a halo. 
Also, the smallest rectangular box enclosing each halo is found.
Next we order the groups according to their 
mass and deploy a bottom--up scheme for identifying which groups
and particles are within more massive structures. 
For each halo in turn (starting with the least massive), 
the remainder of the list
is searched to see if the density peak is within
a box containing a more
massive structure; if such a box is found then the halo is associated 
with the more massive structure.  
If a structure already has associated substructures, they will also
belong to the bigger structure. 
If the density peak is not within any other box, the search is
repeated to check if there is an overlap of the minimum size boxes, 
and if an overlap is found the halo is associated 
with the more massive structure.  
In this way each structure will either belong
uniquely to a combined group or be an isolated structure. We then
calculate the minimal size box of each combined group or isolated
structure and read in all particles inside this box, as long as
they have not already been identified as 
belonging to another structure.  Note that in this way each particle
which has been associated with a structure by FOF
belongs {\em uniquely} to a
family. However a small number of particles which have not been
associated to a structure by FOF (either by being isolated or
belonging to a group with less than 10 particles) might belong to more
than one family; these particles are usually at the margin of the
family and are not significant for the further analysis.

\begin{figure}
\epsfig{file=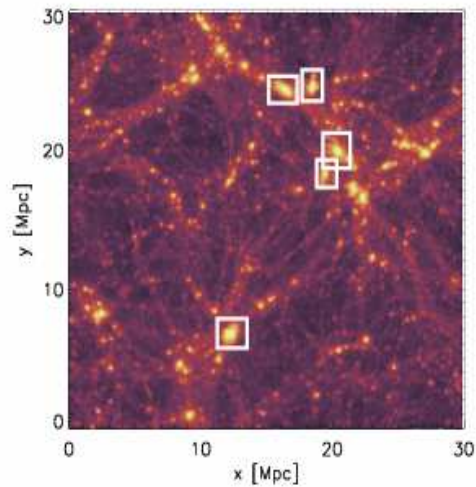,height=7cm,width=7cm}
\caption{Projection of a simulation with $256^3$ particles. The boxes
are the minimum size boxes of the five most massive halos found
with FOF with a linking length of $R_{\rm link} = 1/\left[5\sqrt[3]{\bar
n}\right]$. Note that the apparent overlap of the boxes is just a projection effect.}
\label{fig:sim256}
\end{figure}
In Fig.~\ref{fig:sim256} we show the projection of the simulation with $256^3$
particles in a box of length $L=20 h^{-1} {\rm Mpc}$ at a redshift
$z=1$ and a mass resolution of $\approx 4\times 10^7 {\rm M}_\odot/h$. The boxes are
the minimum size boxes of the five most massive structures in the
simulation. 

This rough analysis of structure enables one already to estimate
the mass distribution in the simulation.
\begin{figure}
\epsfig{file=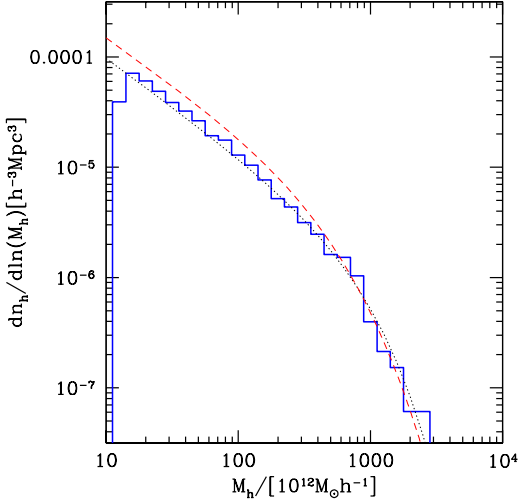,height=7cm,width=7cm}
\caption{Mass function of clusters with more than 5000 particles for
the $1024^3$ simulation (right). The  
  dotted line is the Schechter function  with a slope of
$\alpha =0.9$  and  
exponential cut-off at a scale $M_* =8.0\times 10^{14} h^{-1}
\msun$. For comparison we also plot the mass function from
{\protect\cite{Evrard:2002a}} (dashed line).}  
\label{fig:mass_full}
\end{figure}
In Fig.~\ref{fig:mass_full} we show the mass distribution of families for the
$1024^3$ simulation described in Table \ref{tab:simulations}. 
The dominance of low mass objects is clear. 
Also we show a fit to the slope of the distribution
with to a generalised Schechter function \citep{Press:74a,Schechter:76a}
$dn_h/dln(M_h) = N_*(M_h/M_*)^{-\alpha}\exp(-M_h/M_*)$ and obtain $\alpha
\approx 0.9$. The fit was performed with a nonlinear least-squares
 Marquadt-Levenberg algorithm. Note that at this stage we plot the mass function of the
families, which makes it harder to compare with the standard
Press-Schechter prescription, which assumes virial masses
and does not take into account the
linking of overlapping structures, however our findings are
consistent with previous work \citep{Ghigna:00a}. The dashed line in
Figure \ref{fig:mass_full} from shows the distribution measured by \cite{Evrard:2002a}, which establishes that this rough catalog agrees well with standard expectations.

After this first step we have identified large structures
in the simulation and assigned all particles which potentially belong
to these structures. This will enable us in the next step to refine
the analysis within a single family.

\subsection{Identification of substructure and bound particles in
halos}\label{sec:id}
We are now in the position to study a single family in more detail. We first perform an identification of
groups within one family
using the Denmax algorithm \citep{Bertschinger:91a,Gelb:94a}.
Denmax first interpolates the density field $\rho$ by applying a Gaussian
kernel with a given smoothing length $\rs$ to the particle
positions. The particles are then shifted along the density gradient
via the fluid equation
\beq
	\frac{d{ \mathbf x}}{d\tau} = {\mathbf\nabla} \frac{\delta \rho }{\rho}\, .
\eeq
Each particle moves toward a density maximum where it comes to rest, or
more probably oscillates around the peak. 
The groups are then identified by using the FOF scheme on the
shifted particles, with a linking length comparable to $\rs$. 
We use a much smaller smoothing
length $\rs$ than the linking length $R_{\rm link}$ in the FOF scheme
used previously
for finding the rough structures. We take
\beq
	\rs = f_{\rm sub}\epsilon \, ,
\eeq
where $\epsilon$ is the softening length of the simulation and
$f_{\rm sub}$ is a free parameter in our analysis, which we typically choose 
to be
$f_{\rm sub}=5$. This choice ensures that we identify 
the smallest structures which are still
above the resolution threshold of the simulation
\citep{Ghigna:00a}. We also set the threshold for the minimum number of
particles in a group to 10. In this way we obtain a list of groups
within the single family. 

After the refined Denmax step there are still particles which are not
assigned to any group with more than 10 particles. For each such
particle, we locate the nearest neighbour structures and calculate
the distance $\delta r$ to their density peak positions. We also calculate the
distance to the peak of the most massive group, which we call the {\em
mother halo}. We then calculate $m/\delta r^2$, where $m$ is the mass
of the neighbour, and assign the particle to the group (or the
mother) where this quantity is maximal. We note that any
mis-assignments made at this stage will be rectified at a later stage
in the analysis, and the purpose of this simple criterion is to minimise 
the necessary amount of reassignment.

\begin{figure}
\epsfig{file=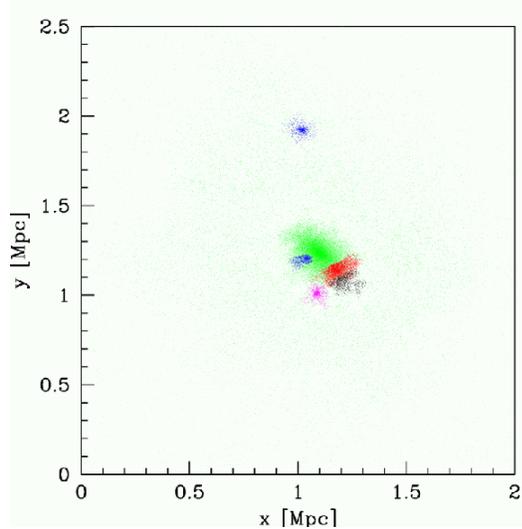,height=7cm,width=7cm}
\caption{Projection of all particles in the mother halo (background), and the five most massive substructures found by the
refined Denmax run. The cluster has a total mass of $1.3\times 10^{13}h^{-1}\msun$ and
the substructures range between $1.2\times10^{9}h^{-1}\msun$ and
$2.3\times10^{11}h^{-1}\msun$ for this simulation.}   
\label{fig:pdenmax}
\end{figure}
As an example, we show
in Fig.~\ref{fig:pdenmax} the five most massive substructures
identified in the most massive mother halo of the $256^3$ simulation,
which has initially $\approx 321,000$ particles, or a mass of $1.3 \times
10^{13}h^{-1}\msun$. The masses of all the substructures vary between
$1.2\times 10^{9}h^{-1}\msun$ and $2.3\times 10^{11} h^{-1} \msun$,
where we assume we can reliably identify a substructure if it comprises
of at least 30 particles.

The next step is the build up of the family tree within this
family. In order to obtain the family tree, we calculate the
minimum size box which contains each identified substructure.
Then, as before, we apply a bottom--up scheme starting with the lowest
mass halo and determine if its density peak is within the minimal
box enclosing a more massive structure. The structure with the
lowest mass which contains the halo is identified to be the
{\em mother} of this halo, while the halo becomes the {\em daughter}
and hence a substructure of the mother. If the density peak is not
within any other halo, we check if the minimal box is overlapping with
any other box. In this case we take the lowest mass overlap halo as
the mother. 
We then move to the next more massive halo
and repeat the procedure. Once we have identified the mother, all the
substructures of the daughter will also become daughters 
of the mother. In this way we obtain a unique mother for each halo, and
for each mother a list of daughters which contains 
all substructures of the hierarchy. We should actually talk of
daughters, grand-daughters, great grand-daughters and so on, but there
is no need to distinguish daughters and grand-daughters from a
mother's point of view, as long as each daughter knows who her mother
is--- which is ensured by our procedure. In other words, each mother knows
about the whole younger generation, but only her mother from among her 
ancestors. ``Isolated'' substructures will have the original mother
halo as a mother. 
\begin{figure}
\epsfig{file=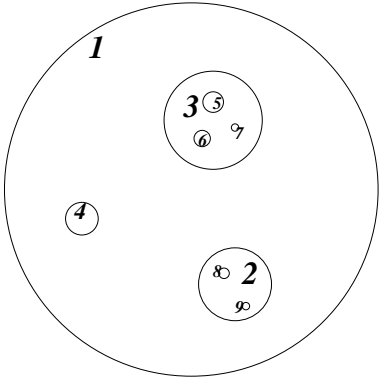,height=5cm,width=5cm}
\caption{Schematic description of the family tree. 
Halo number 1 counts all subhalos 2 through 9 as her daughters. 
Halo 1 is the mother of subhalos 2, 3, and 4.
Subhalos 5, 6, and 7 are also the daughters of subhalo 3,
and see 3 as their mother; 7 and 8 are in a similar relation to 
subhalo 2.  Number 4 is an isolated subhalo. }
\label{fig:famtree}
\end{figure}
In Fig.~\ref{fig:famtree} we show schematically the build up of a
family tree.

We further introduce a threshold particle number $N_t$. Structures
with fewer particles than $N_t$ are dissolved into their associated mothers. Typically we choose
$N_t=30$, discarding smaller groups found earlier. In
\cite{Ghigna:00a} a threshold of $N_t=16$ has been used for using
halos as tracers, but $N_t = 32$ for the reliable analysis of
properties of halos. However \cite{Diemand:04b} find that the results
are only stable for $N_t = 100$. This is necessary to resolve a
complete sample of subhalos. 
\begin{figure*}
\hrule{\epsfig{file=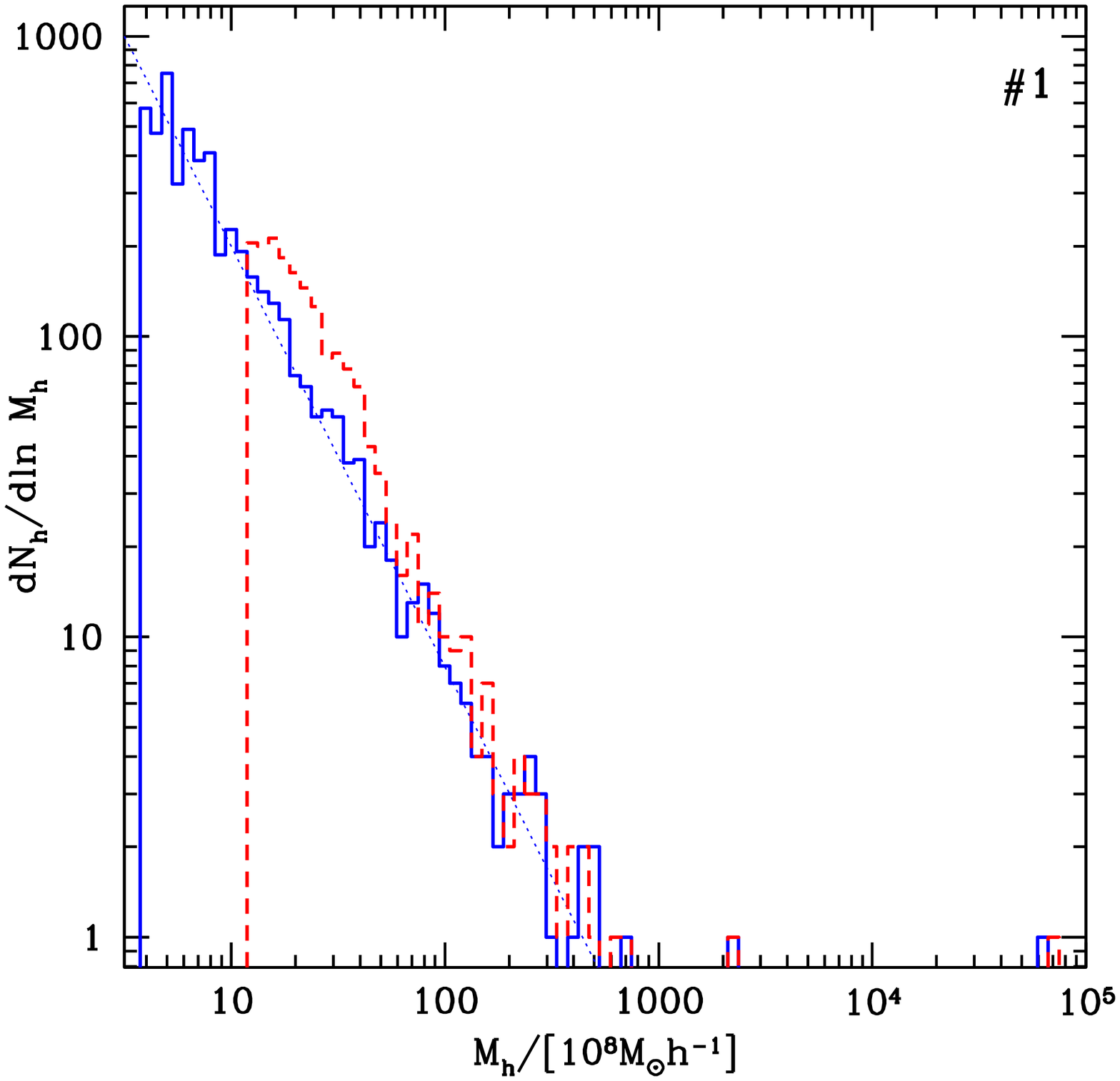,height=7cm,width=7cm}\epsfig{file=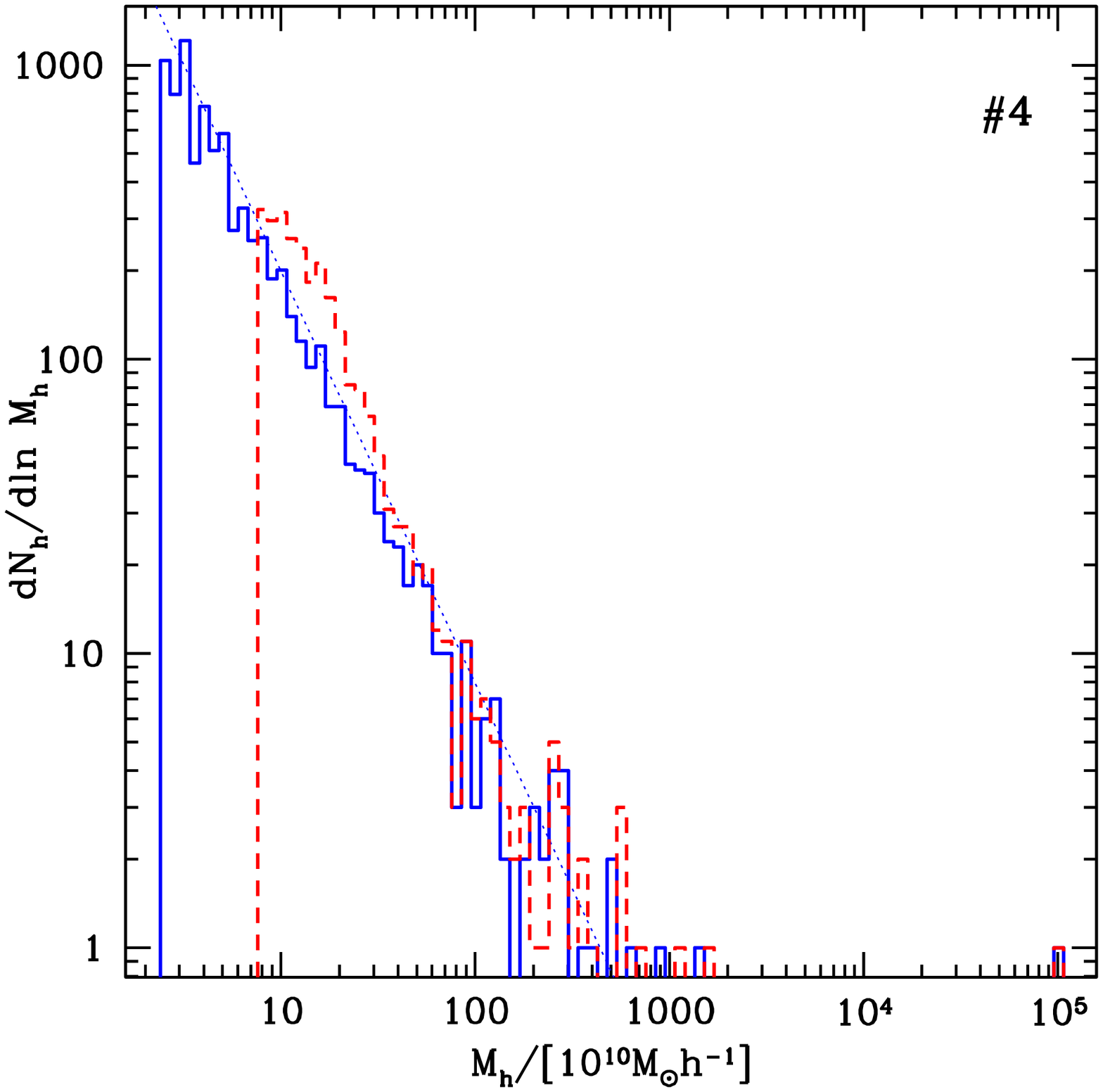,height=7cm,width=7cm}}
\caption{{\it Left:} Mass function of the substructures of
the most massive halo of the $256^3$ simulation, before (solid line) and
after (dashed line) 
structures with less than $N_{\rm t}=30$ have been dissolved
into their mothers.  The mother halo has $\approx 160,000$
particles or a mass of $6.5\times 10^{12}h^{-1}\msun$. The dotted line
is from a power law $dN_h/dln(M_h)\propto M_h^{-1.4}$.  {\it Right:}  The
same plot for the 4th most massive halo of the $1024^3$ simulation
with $\approx 392,000$ particles and a mass
of $1.1\times 10^{15}h^{-1} \msun$ in the mother halo.}
\label{fig:mass_denmax}
\end{figure*}
In Fig.~\ref{fig:mass_denmax} we show the distribution of
substructures masses in the most massive halo of the $256^3$ run, 
and in the 4th most massive halo from the $1024^3$ simulation. In the
following we call these halos cluster \#1 (left) and cluster \#4 (right). The
entire structure, including the mother halo and all daughters, has a mass of
$1.3 \times 10^{13}h^{-1} \msun$ for cluster \#1 and
$1.6 \times 10^{15}h^{-1} \msun$ for cluster \#4. We
clearly see 
the large number of small structures (solid lines); the original
distribution follows roughly a scaling of $dN_h/dln(M_h) \propto M_h^{-1.4}$ (dotted
lines). This becomes even steeper if the halos with less than $N_{\rm t}=30$
particles are dissolved into their mothers (dashed lines). The Denmax
routine had originally recovered about $5100$ substructures 
which have more than ten particles within cluster \#1, 
and $7760$ halos within cluster \#4. About 20-25\% of the total
mass of the structure is in halos with less than $30$
particles as identified by Denmax. After halos with less than 
$30$ particles have been dissolved into their mothers we are left
with about $1790$ substructures in cluster \#1 and 2590 in cluster
\#4. 
During this procedure the original mother halo gained 
$5.7\times 10^{11}h^{-1}\msun$ for cluster \#1
and $6.4\times 10^{13}h^{-1} \msun$ for halo \#4. The rest of the mass is distributed among
the lower mass halos, as seen in the dashed histograms in Fig.~\ref{fig:mass_denmax}.

As mentioned in the introduction, Denmax itself has been
applied in a hierarchical way, either as part of SKID by applying three
different smoothing lengths $l_{\rm link} = 1.5,5,10 l_{\rm soft}$
\citep{Ghigna:00a}, or by using it on larger scales
 with $\rs = 0.2 {\bar n}^{-1/3}$ and re-analysing each
halo with $\rs = 0.1 {\bar n}^{-1/3}$ \citep{Neyrinck:03a}. 
The reason for this is that in general there is
no single smoothing length which is suited to find structures over a
large mass range in the simulation. If the smoothing length is too
large then small structures are not resolved, and if it is too small
then large structures are broken up. 
We choose a small smoothing length
and recombine larger objects using the family tree hierarchy.

We have now a clearly defined, geometrically based picture of
substructures, which we can proceed to analyse in a more physical
fashion so that unbound particles are culled out.
In some situations the Denmax procedure may err in assigning
some particles to substructures.  Imagine a particle which is
dynamically a part of the mother halo:  the Denmax algorithm will 
move this particle toward the cluster centre, but if 
a significant substructure just happens to intervene, the particle
will reach this local maximum and stop.
Thus there will be particles extending in a radial wedge
outside of any bound structure arbitrarily attached to it, even if
they are gravitationally not bound to it.
To correct for such unphysical identifications,
we need now a post--identification dynamical treatment of the halos.

\subsubsection{Velocity outliers}
It can be shown \citep{Binney:87a} that
the rms escape velocity from a finite, bound self-gravitating system 
is related to the rms velocity by $\left<v^2_{\rm esc}\right>=4 v_{\rm
rms}^2$. Thus particles having a velocity greater than $\sqrt{2\left<v_{\rm
esc}^2\right>}=\sqrt{8}v_{\rm rms}$ are very unlikely to be bound to
the structure. One way to calculate the escape velocity is by
measuring the maximum value of the circular velocity $v_{\rm circ}(r) =
\sqrt{GM(r)/r}$; by assuming an NFW profile this can be related to
the escape velocity \citep{Klypin:99a}. However this method relies on
the NFW profile which we do {\em not} want to assume at this stage.

To remove unbound particles from a substructure, 
we will instead proceed with a first approximation 
by calculating the typical rms velocity and removing
particles which are statistical outliers.  But we cannot calculate the
velocity dispersion until we know the
true centre of mass (CoM) velocity, so--- because we have not removed unbound
particles from the structure--- we must proceed iteratively, beginning
with an approximation for the
CoM.  We choose the density peak of the substructure (not
including its daughters) as a first approximation to the CoM. In order to
obtain the CoM velocity we calculate the median of the velocity of the
$N_v=100$ nearest neighbours to the density peak within the
structure. If the number of particles is less than 30, we take half
the particles of the structure. In order to obtain a valid answer we must pay attention
to binaries, which could bias the result to large velocities. Hence we
identify binaries by searching the whole simulation for bound
pairs. We so far have not found bound pairs of particles in all the
simulations we studied, which also provides evidence that the
simulation is not over-resolved.  If we did find a bound pair, the
two particles would be replaced by a single particle with twice the mass,
and the CoM position and CoM velocity of the pair. This ensures that
we do not encounter velocity biases due to binaries. We then can
proceed to calculate the rms velocity 
$v_{\rm rms}^2=
\left<\left({\mathbf v}_{\rm part}-{\mathbf v}_{\rm cm}\right)^2\right>$
for the $N_v=100$ particles around the density peak.
All particles in the substructure which have a velocity  
\beq
	{\mathbf v}_{\rm part}^2 >f_{{\rm cut}}v_{\rm rms}^2\, ,
\eeq
are then removed and added to their associated mother structure.
We iterate this process until the mass change of the substructure is
less then $5\%$. We perform this velocity cut at two levels: first
we use $f_{{\rm cut}}=8$ as noted earlier, and as mentioned above $N_v=100$ particles
for the CoM velocity and rms velocity calculation. Then choosing a
tighter limit with $f_{\rm cut}=6$, we find the centre of mass mean
velocity and velocity dispersion of the inner half of the particles
and repeat the process.

\begin{figure*}
\hrule{\epsfig{file=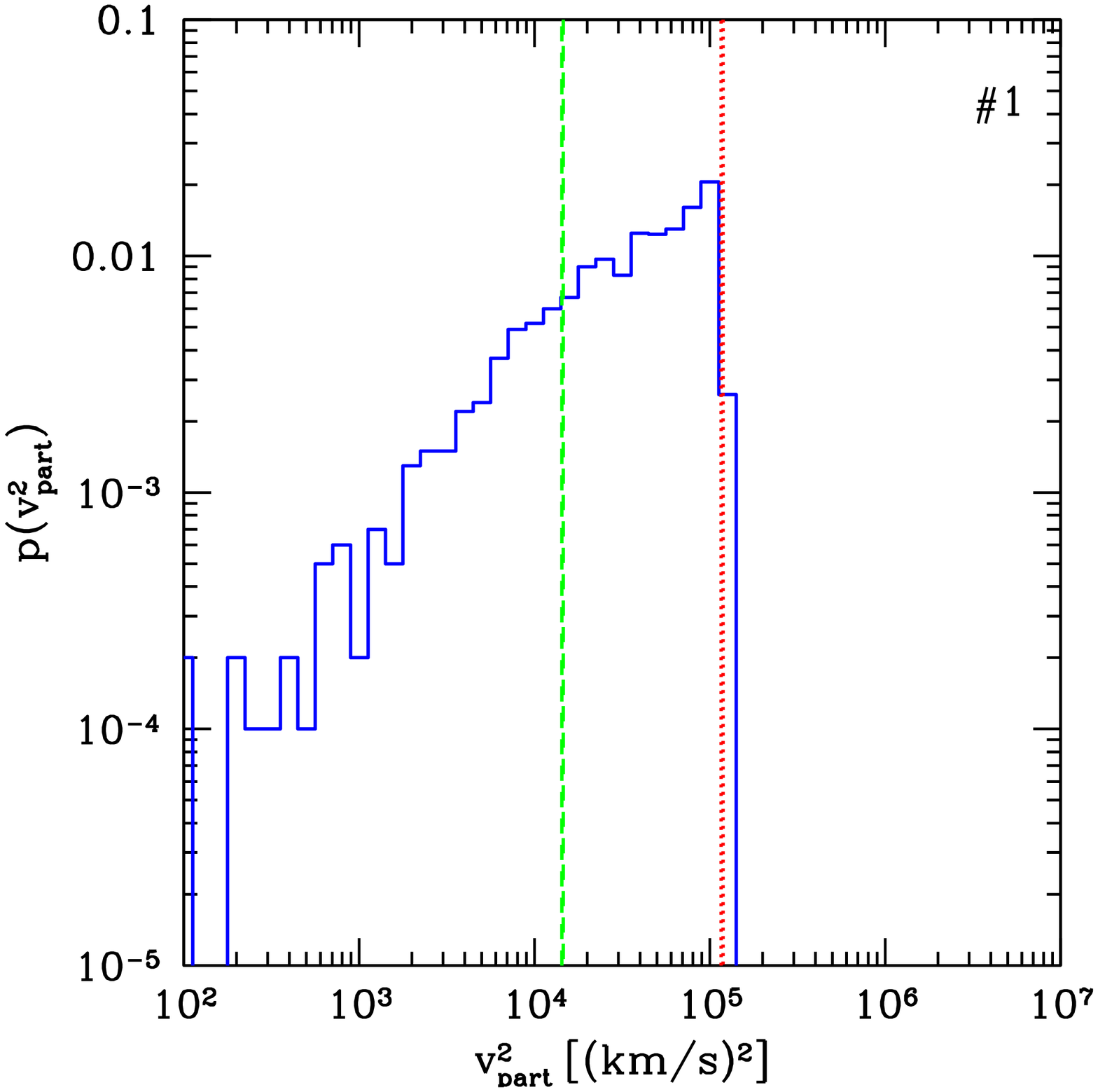,height=7cm,width=7cm}\epsfig{file=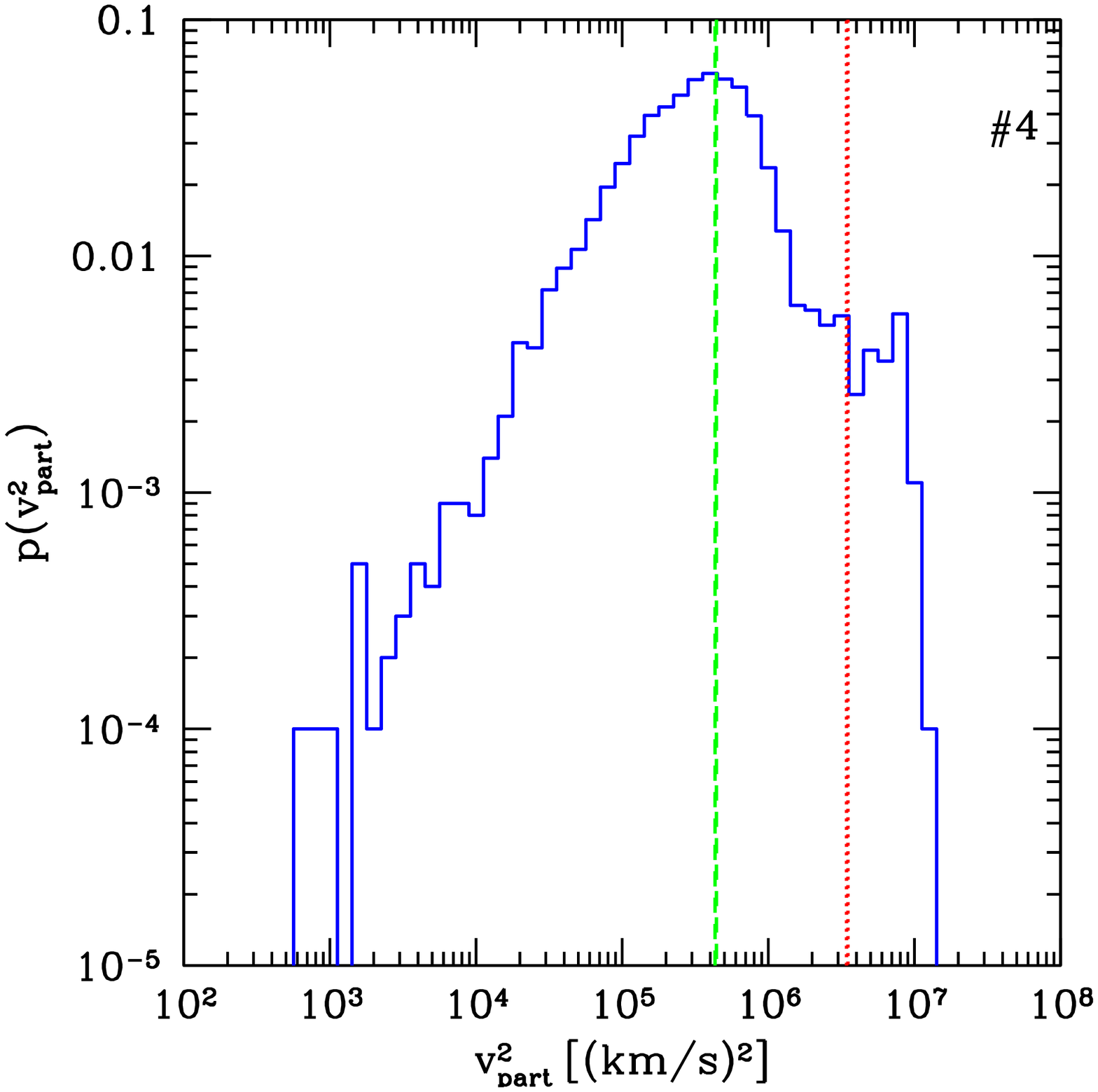,height=7cm,width=7cm}}
\caption{Distribution of the particle velocities in the most massive substructure, the velocity
threshold $8v_{\rm rms}^2$ (dotted) and the rms velocity (dashed). All the particles above the
threshold are moved to the associated mother. 
Cluster \#1 is on the left, and \#4 on the right.
} 
\label{fig:vcut}
\end{figure*}
In Fig.~\ref{fig:vcut} we show the velocity distributions of the
particles in the most massive substructure (solid line) in clusters
\#1 and \#4. We also show
the threshold rms velocity (dotted line) during the first step of the
iteration. All particles above this 
threshold are moved to the associated mother. The mass of the mother
halo at the end of this procedure increases just by $4.2\times
10^{11}h^{-1}\msun$ for cluster \#1 and $2.0\times 10^{13}h^{-1}
\msun$ for cluster \#4, where most of the change occurs during the first cut-off
scheme. After the removal of the velocity outliers we again dissolve
halos with less than $N_t=30$ particles into their mothers. 

At the end of this step we recalculate the CoM and $v_{\rm rms}$ for
the subhalo and
then  we move daughters which have a
faster CoM velocity than $\sqrt{6}v_{\rm  rms}$ to the associated mother
of the substructure under consideration. 

\subsubsection{Tree calculation of potentials and bound particles}\label{sec:unbind}
We now reach the step where we can remove particles which have a total
energy larger than zero in the centre of mass frame of
the structure to which they belong. We will check within each
substructure which particles are bound  
to it. We calculate the CoM of a substructure including {\em
all} its daughters and compute the potential $\phi$ of the
particles within the substructure.
The potential calculation is done using an adaption of a tree code by
Hernquist (1987). Note that we switch to an exact direct summation of
of the potential energy if there are less than 100 particles in the
system. 
The total energy of a particle is then
\beq
	E_{\rm tot} = m\phi+\frac{1}{2}m\left({\mathbf v}_{\rm
part}-{\mathbf v}_{\rm cm}\right)^2 \; ,
\eeq
where $m$ is the mass of a particle, $\phi$ the potential from all
the other masses within the substructure, and ${\mathbf
v}_{\rm cm}$ the CoM 
velocity of the substructure. We calculate $E_{\rm tot}$ for each particle
and then remove the third of the unbound particles with the highest
energies,  moving them to their associated mother
structure.  Note that we choose only a third of the particles because
otherwise particles are removed to quickly without taking into account
that the CoM velocity, and hence the kinetic energy, is changing with
each removed particle. Ideally one should remove only one particle at
a time, as it is done in SKID \citep{Stadel:97a}, but this is too time
consuming for hundreds of halos with over $10^5$ particles. We tested
different fractions and observed that one third was the largest number
which results in a stable result. 
We then recalculate the CoM and iterate this step until there is no
change in the mass of the system.

\begin{figure*}
\hrule{\epsfig{file=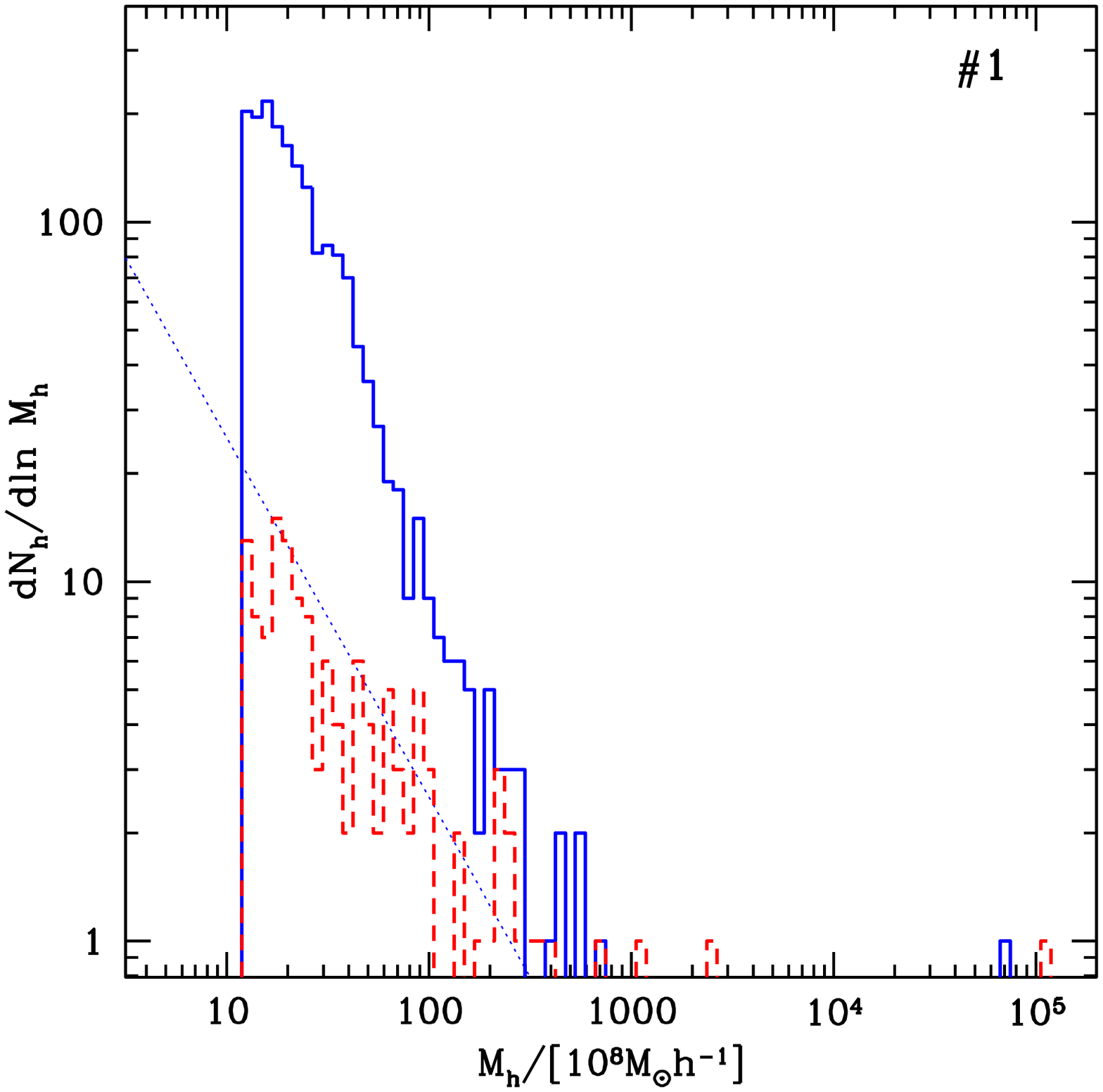,height=7cm,width=7cm}\epsfig{file=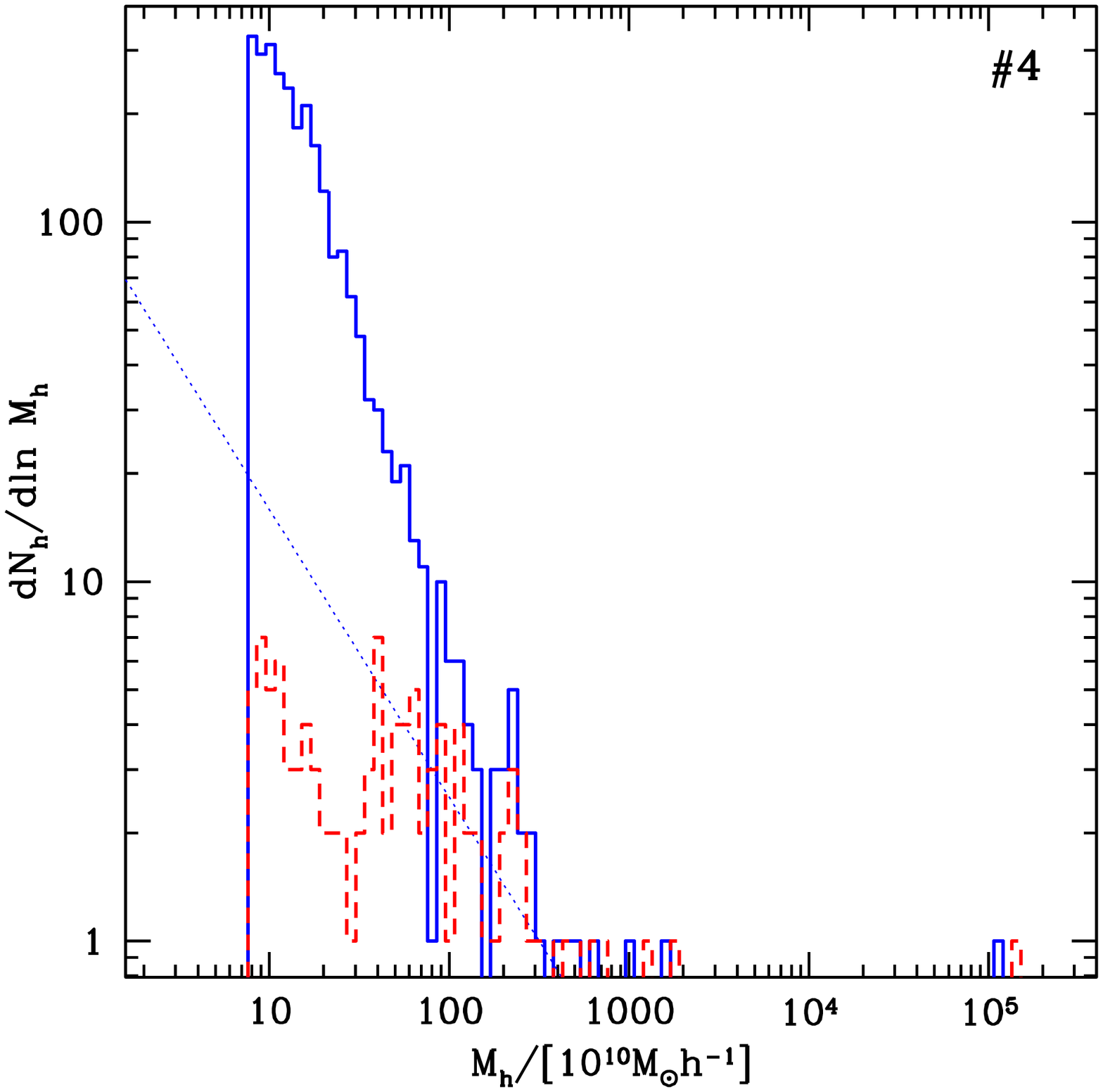,height=7cm,width=7cm}}
\caption{Mass function of substructure in the test clusters \#1 (left)
and \#4 (right) after
particles with high velocities have been moved to the associated
mothers (solid line) and after unbound particles have been moved to
the mother structure (dashed line). Note that the power laws are now
$dN_h/dln(M_h) \propto M_h^{-1.0}$ for \#1 and $dN_h/dln(M_h) \propto
M_h^{-0.8}$ for \#4.}
\label{fig:bound}
\end{figure*}
In Fig.~\ref{fig:bound} we show the mass distribution before and after
unbound particles have been moved to the mother structures. Note that all
daughters with less than $N_{\rm t}=30$ particles have been dissolved
into their mothers. There are many unbound particles in the
substructures which returned to the original mother. 
The mother in halo \#1 now has a mass of
$1.2\times 10^{13}h^{-1} \msun$ which corresponds to $\approx 293,000$
particles; there are now only $134$ daughters with a
total mass of $1.1\times 10^{12} h^{-1} \msun$. 
The mother of halo \#4 has a mass of $1.5\times 10^{15} h^{-1}\msun$ or 
$582,000$ particles,  with
$1.2\times 10^{14} h^{-1} \msun$ remaining in $106$ daughters. 

Before we proceed with the next step
we will remove any daughter which is not bound to its mother. We
approximate the potential energy for the daughters by 
\beq
	E_{\rm pot}^{\rm d} = -G\frac{m_{\rm d}M(r_{\rm d})}{r_{\rm
d}}-G\int_{r_d}^\infty \frac{m_{\rm d}}{r}\; dM(r)   \, ,
\eeq
where $m_{\rm d}$ is the daughter mass,
$r_{\rm d}$ is the distance of the CoM of the daughter to the
CoM of its mother, and $M(r)$ is the total mass of the
mother within radius $r$ including all other daughters. We then
can calculate the kinetic energy of the daughter with respect to the
centre of mass its mother. If the daughter is {\em not} bound to
her mother we move her to the mother of the mother.

\subsubsection{Search for Hyper-structures}
In order to obtain a stable algorithm with respect to the
smoothing length for the refined Denmax procedure, we need, as
noted above, to
look for ``hyper-structures'', groups of substructures which are
gravitationally collectively bound to one another.

This problem has been addressed previously by combining the
SKID algorithm with an adaptive FOF analysis \citep{Diemand:04b};
we will take a different
approach here.
In order to do this we investigate primary substructures,
ie. structures which are {\em direct} daughters of the largest
structure which is the mother structure. 
For each such primary substructure, we calculate
the distance $\delta r_i$ to each other primary substructure with
mass $m_i$, and examine the one with the maximal $m_i/\delta r_i^3$
as follows;  note that the masses include all daughters
of the primary substructures.
If these two
structures are bound with respect to their common CoM, they form a
hyper-structure;  the less massive of the two becomes a daughter of
the more massive structure. We than re-calculate the CoM and the
maximum extension box of the new hyper-structure, and check each
particle of the mother within this box.  If it is bound to the
hyper-structure, we then move it from the mother to this
hyper-structure. We will iterate this step three times.
Note that for both halos the mass in the mother structure does not
change significantly during this step.

In this fashion bound objects, whose identity
is independent of the geometrical tool used to analyse substructure,
are assembled.

\subsubsection{Final Steps: Daughters and Particles unbound to Entire Family}
The next step we perform is to remove daughters which are not
bound to the biggest structure, the mother halo. And finally we
remove particles which are not bound to the family tree at all. 
For both halos none of the daughters is unbound and the number of unbound
particles is negligible.

\begin{figure*}
\hrule{\epsfig{file=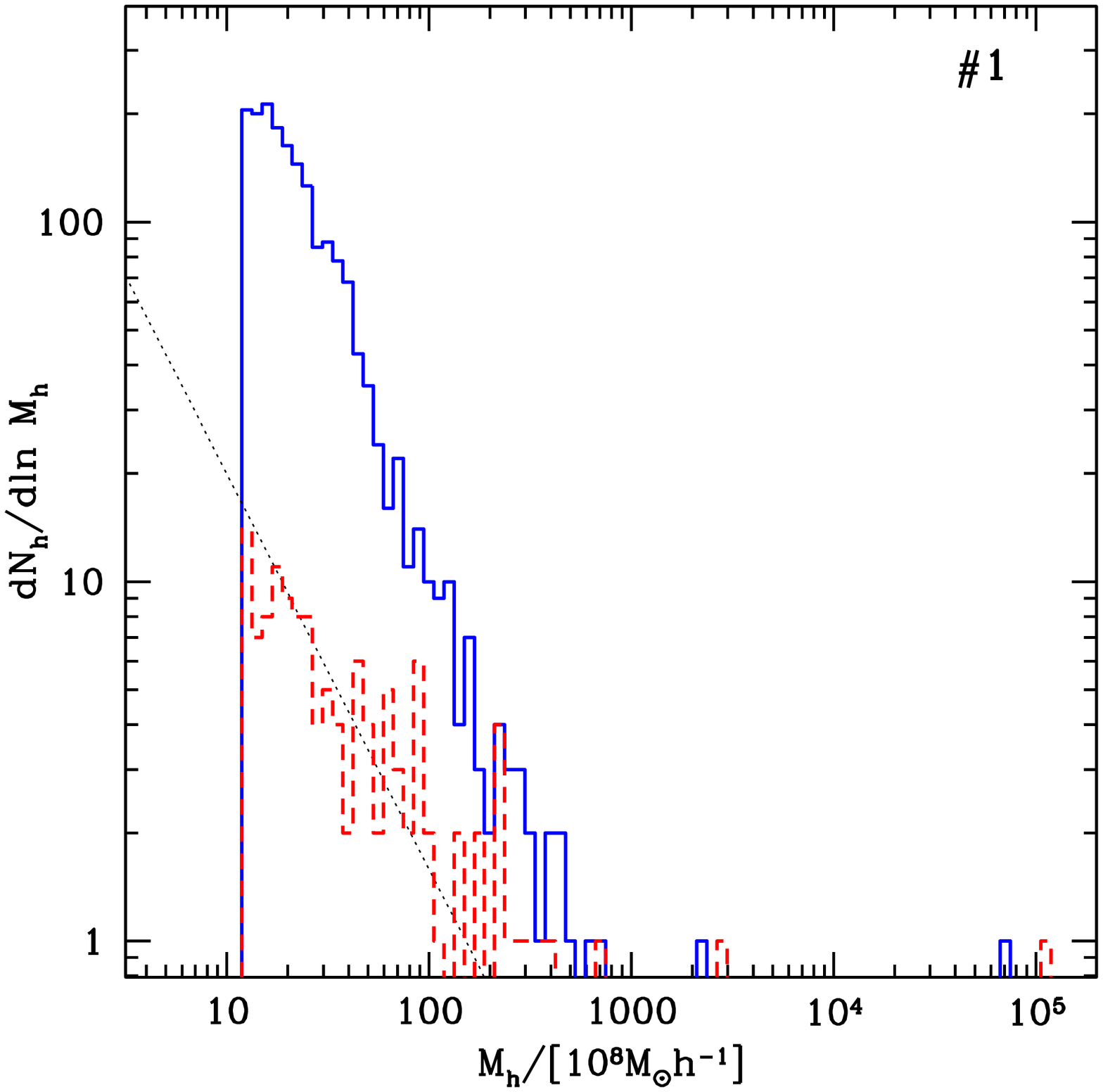,height=7cm,width=7cm}\epsfig{file=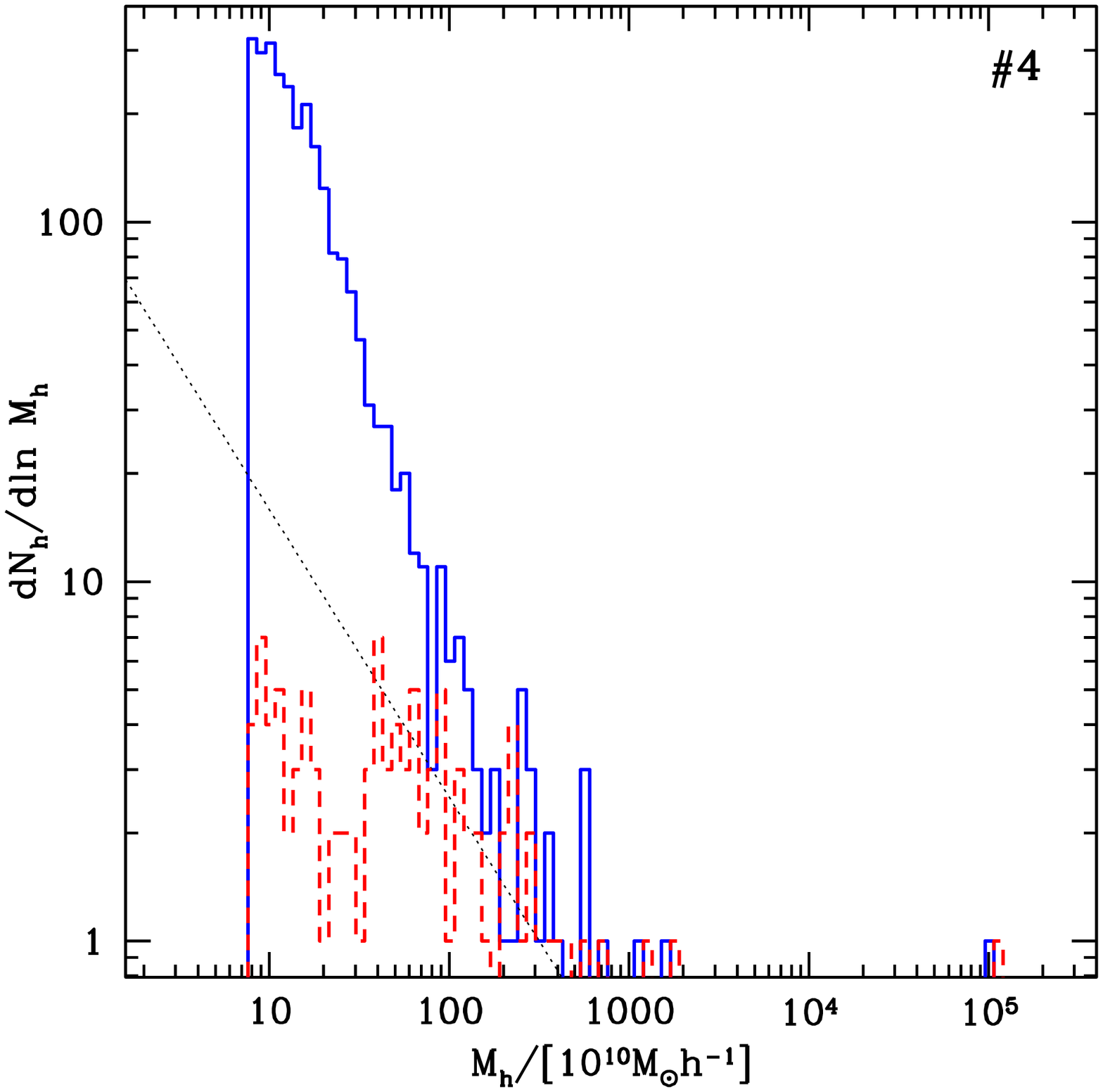,height=7cm,width=7cm}}
\caption{Mass function of substructures in the halo after the
refined Denmax run and the mass cut (solid), and at the end of the
the iterative scheme (dashed). The dotted lines show the power laws $dN_h/dln(M_h)
\propto M_h^{-1.1}$ (left) and $M_h^{-0.8}$ (right).}
\label{fig:mtot}
\end{figure*}

In Fig.~\ref{fig:mtot} we show the mass distribution from the refined
Denmax run (solid line) and after the unbinding steps (dashed line).  
There
remains  8.8\% of the mass in substructures for \#1 (left) and 
7.4\% for halo \#4 (right). For cluster \#1 
$dN_h/dln(M_h) \propto M_h^{-1.1}$ for small mass halos, 
the distribution in cluster \#4 is roughly approximated by $dN_h/dln(M_h) \propto M_h^{-0.8}$. Note these are only rough power laws.

\subsubsection{Truncation of Halo at the ``virial radius'' and Identification of Companions}
Now we check if we have artificially linked together separate structures
which are only weakly coupled together gravitationally, and if we have
artificially included distant in-falling matter. 
For a $\Lambda$CDM cosmology, it is conventional to define 
the virial mass $M_{\rm vir}$ and radius $R_{\rm vir}$ with
\beq
M_{\rm vir}= \frac{4}{3}\pi R_{\rm vir}^3 \Delta_c(z) \rho_c(z)\, ,
\eeq
where $\rho_c$ is the critical density of the universe, and
the mean over-density $\Delta_c=178\Omega_{\rm m}(z)^{0.45}$.
Thus we make a rank
ordered list of our mother halos, and in each one we start at the
density maximum and proceed outward until we reach the virial radius,
within which the mean over-density is $\Delta_c$.
We 
truncate the halo at this point, removing all particles from outside
the virial radius of the halo
and identifying daughters with centres outside this radius as separate companion
structures. In this
step the mass of the mother halo in cluster \#1 stays almost constant at $1.1
\times 10^{13} h^{-1} \msun$ while 96 subhalos with a total mass of $9.4\times 10^{11} h^{-1}
\msun$ remain. In cluster \#4 the mother mass is reduced to $1.2\times 10^{15} h^{-1} \msun$ with 65 remaining subhalos of a mass of $8.2 \times 10^{13} h^{-1} \msun$. 

\begin{figure*}
\hrule{\epsfig{file=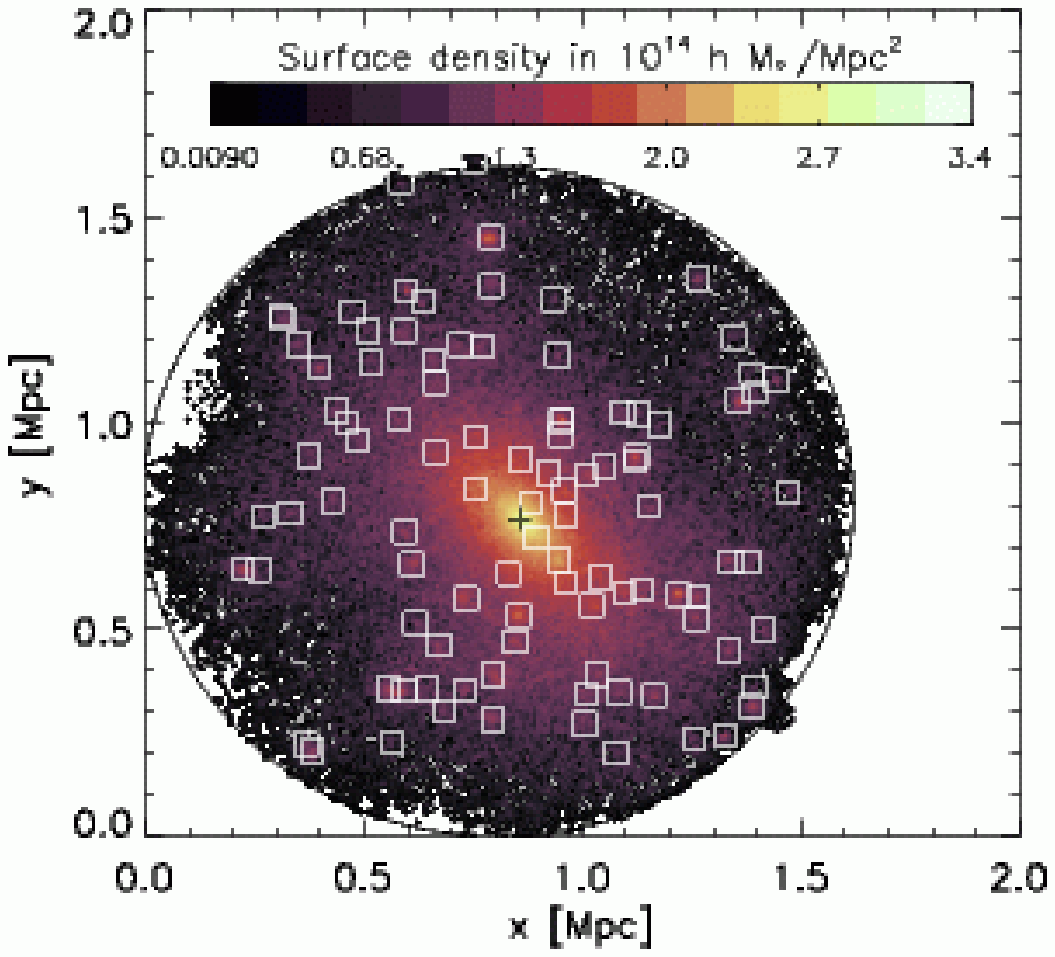,height=6cm,width=6.5cm}\epsfig{file=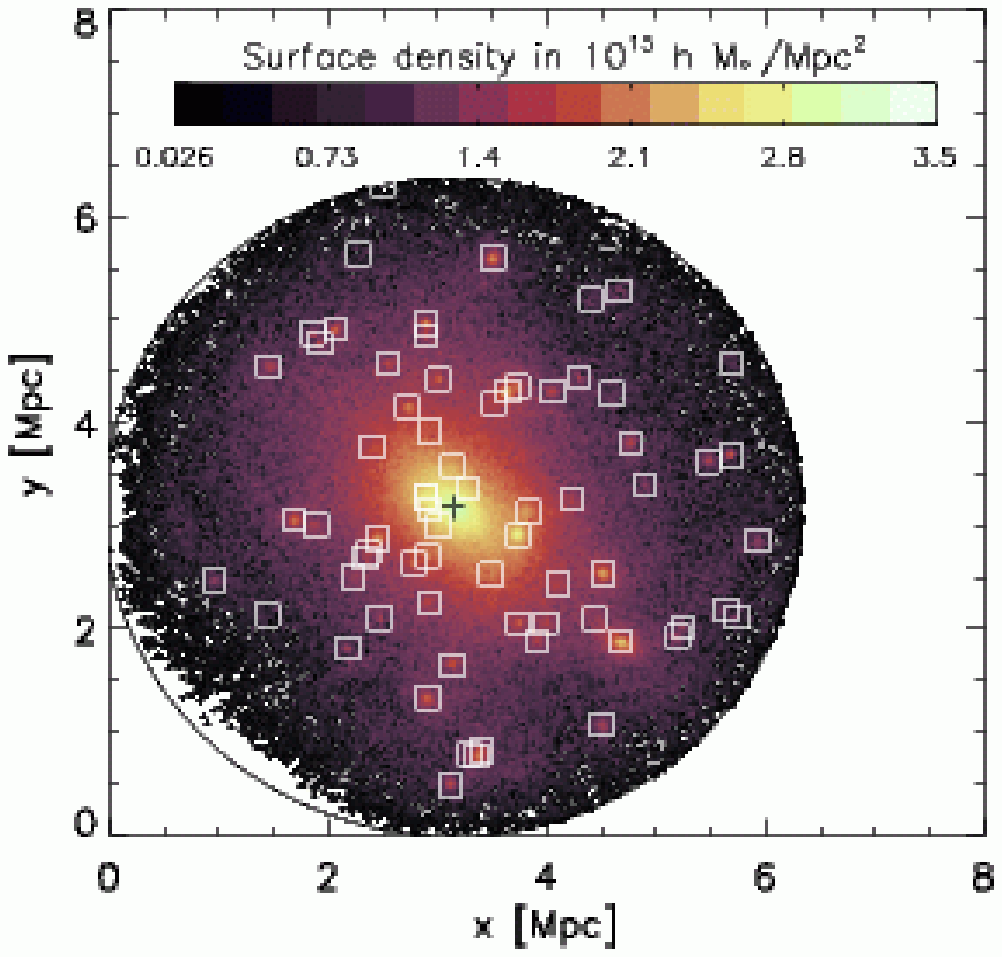,height=6cm,width=6.5cm}}  
\caption{Particles and daughters in halo \# 1 (left) and halo \# 4
(right) after the cleanup
procedure and removal of
particles and subhalos outside the virial radius. The dark
cross marks the centre of the halo and the boxes the
identified and bound daughters. The colour  
corresponds to the surface density as indicated by the
colour bar. Cluster \#1 has 96 subhalos and \#4 has 65.}
\label{fig:haloproj}
\end{figure*}
In Fig. \ref{fig:haloproj} we show the projected density inside the
virial radius of cluster \#1 (left) and cluster \#4 (right), with the daughters
marked. Note that cluster \#1 is at a redshift of $z=1$ while cluster
\#4 at redshift $z=0.05$. Halo \#1 has a virial radius of $r_{\rm
  vir}\approx 0.81\,{\rm Mpc}$ and halo \#4 a radius of $r_{\rm vir} \approx
3.0\, {\rm Mpc}$. Note that for halo \#4 the substructure is much more
centered around the core than for halo \#1. 

\section{Dependence on Parameters and Test of stability}\label{sec:stability}
In this section we discuss the stability of our method, as
different possible choices of the parameters and procedures
may affect the results.
\subsection{Group Finding}
We will first vary the linking length in our rough initial FOF
analysis in order to establish how sensitive we are to this parameter
choice. We perform an analysis with a linking length of $R_{\rm link}
= 0.167{\bar n}^{-1/3}$;  this could potentially lead to a larger
fragmentation of initial halos and families and hence potentially
change our results. We find that the results of this run are almost
identical with results obtained with the original linking
length. For halo \#4 we have after the initial 
fine Denmax run $35\%$ of the mass in substructures (compared to 34\%
in the original run)
which is lowered to $9\%$ (8\%) after we test for bound particles;
after the final virial cut, $8\%$ of the halo
mass in substructures while the original run resulted in slightly lower than $8\%$. This
is due to the fact that our family tree procedure followed by a
refined Denmax run produces almost the same large structures.
 
We did a further consistency check where instead of FOF
we used a rough Denmax run
with a smoothing length of $\rs = 1/5 {\bar n}^{-1/3}$ 
to identify the initial halo list.
The results were
essentially the same as in the original runs.
Hence
we conclude that our method is stable with respect to sensible changes
in the initial halo finding algorithm to within $\pm 1\%$, which is
well below the statistical fluctuation of the sample.

The next halo finding step we perform is the refined Denmax run. We
crucially chose in this step the smallest sensible smoothing length
and then built up the halo hierarchy by our family tree algorithm.
Making the smoothing length smaller than $\rs = 5 \epsilon$
would enter the regime dominated by uncertainties in the force
softening, so we do not extend a stability test in this direction.

Instead we repeated the analysis
with a smoothing
length of $\rs = 10 \epsilon$.  Due to the larger smoothing length, we
find $56\%$ of the mass in substructures after the initial denmax
step for halo \#4; however, when we test for particles which are actually bound to
these structures we obtain already $9\%$ of the mass in
substructures.  After the inclusion of hyper-structures and the density
cut this drops to 7\%, which  is in excellent agreement compared to
the run with a smoothing length $\rs = 5\epsilon$ ($8\%$).

We hence conclude that we have a
reasonably stable criterion if the smoothing length is chosen within a
reasonable range. Of course, as the smoothing length is made larger
we will miss more and more structures.

\subsection{Removal of unbound particles}
The first step of removing unbound particles is performed by removing
velocity outliers in a gentle way. Since we do this already in two
steps with first a gentler and then a harder cut-off at $8v_{\rm rms}^2$
and $6v_{\rm rms}^2$ we established that most of the cut is happening
during the first iteration step. However the velocity cut does not
change the mass fraction significantly. Final results do not depend on
the specific numbers $(8,6) \times v_{\rm rms}^2$, as long as
we approach the final cut gradually. Furthermore,
we note that this cut was
mainly done to avoid an unphysical bias toward large CoM velocities,
which is important for the calculation of the kinetic energies with
respect to the centre of mass. 

\subsection{Virial Cut}\label{sec:halosize}
Since the definition of a mass or size of a halo is to some extent
arbitrary (see for example:
\cite{Jenkins:01a,Evrard:2002a,White:02a}), 
we will investigate how this
definition influences our results.  We chose initially the virial mass
and radius corresponding in a $\Lambda$CDM cosmology to the over-density 
$\Delta_c(z)=178\Omega_{\rm m}(z)^{0.45}$.
We saw already in the
comparison of the $256^3$ and $1024^3$ simulations that, after this density cut, the
fraction of mass in substructures can be quite different. However,
the simulation with $256^3$ was also at a redshift of $z=1$, compared
to $z=0$ for the $1024^3$ simulation. Hence we performed an analysis
of the the $1024^3$ run where we chose the cut-off over-density to be
$\Delta_c = 200$ in agreement with another commonly used
definition.  With this cut-off the final mass-fraction in
subhalos only decreases from $8\%$ to $7\%$ for halo $\# 4$.

\section{Application to a Different Simulation}
In order to test our algorithm we applied it to a simulation provided
by \citet{Diemand:04b}. We chose their cluster $D6$, where the
simulation was done with a smoothing length of $\epsilon = 3.6\,{\rm
  kpc}$, which is comparable to our runs. \citet{Diemand:04b} state
that their halo finding is complete for halos with more than 100
particles and they find about 5\% of the mass in substructures. 
We also obtain with our scheme 5\% of the mass in substructures, which
is excellent agreement given the difference of the analysis
methods. \citet{Diemand:04b} use 
a hierarchical version of the SKID algorithm which is based on
DENMAX. They perform the unbinding iteratively and exactly with no
approximation like the one discussed in Section \ref{sec:id}. 

In order to get a further inside into the statistics of
substructures we compare the cumulative massfunctions of halo $D6$ from our analysis and the analysis by \citet{Diemand:04b}.
\begin{figure}
\epsfig{file=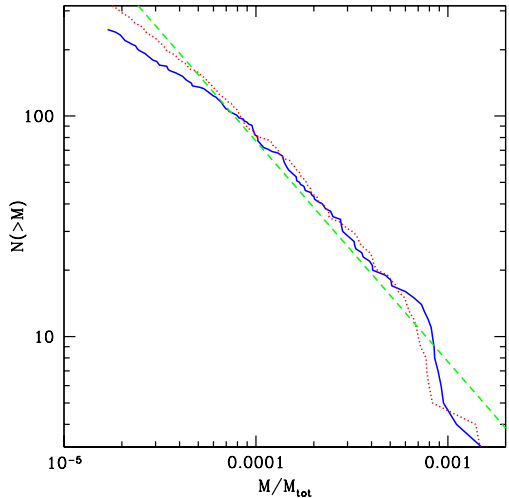,height=7cm,width=7cm}
\caption{Cumulative mass function for substructures with more than
  $30$ particles in halo $D6$ as
  presented by \citet{Diemand:04b}. The solid line is from the
  algorithm presented here, the dotted line from \citep{Diemand:04b}
  and the dashed line has a slope of $M^{-1}$.}
\label{fig:cummass}
\end{figure}
In Fig.~\ref{fig:cummass} we show the cumulative mass function for
substructures in the test halo $D6$. The solid line is from our
analysis and the dashed line from \citep{Diemand:04b}. They look both
very similar and scale very closely to $M^{-1}$ until a cut-off at
less than a thousandth of the total cluster mass. Our analysis results
in slightly more substructures than the one of \citep{Diemand:04b}. We
find 272 substructure while they find 241. This is actually strikingly
similar given the difference of the presented algorithms and the
overall mass fraction in substructures for both analysis is at the 5\%
level within 1\% uncertainty.

\section{Conclusion}
In this paper we have established a fast and stable algorithm to
identify vast numbers of substructures in large N-body
simulations in a speedy fashion. For example to analyse the most
massive halo of about 1.5 million particles takes about 8 hours on
a SUN BLADE 2000 on a single 900 MHz processor with 3 Gbyte RAM. We
established an approximate method to identify and remove unbound
particles from subhalos, which allows for the efficient
calculation of bound structures. We analysed three simulations, two
done by the TPM code developed by \citet{Bode:00a} and one by \citet{Diemand:04b}. For
all three we find similar mass fractions of about
5-8\%. \citep{Diemand:04b} find about 5\% of the mass in
substructures, which is identical with our findings.

The fraction of substructure in a cold dark matter cluster is to some
extend a question of definition. If one for instance is interested in
strong lensing, which tests the distribution of matter, the question
of bound or unbound structures is irrelevant. However, if substructures
are the places where galaxies form, a full dynamical treatment is
relevant and requires the inclusion of {\em all} forces, including the
ones from internal and external potential.

To conclude we emphasize that the presented algorithm is stable and
fast and ready to be employed for large cosmological data sets as well
as detailed simulations of clusters of galaxies.

\section*{Acknowledgment}
We thank A.~Amara, J.~Diemand, G.~Efstathiou, S.~Kazantzidis, A.~Kravtsov, B.~Moore and T.~Naab
for useful discussions, B.~Moore and J.~Diemand for the provision of the external simulations and the referee for valuable suggestions. The parallel computations were done in part at
the UK National Cosmology Supercomputer Center funded by 
PPARC, HEFCE and Silicon Graphics / Cray Research. 
This research was supported by the National Computational Science
Alliance under NSF Cooperative Agreement ASC97-40300, PACI Subaward 766;
also by NASA/GSFC (NAG5-9284).  Computer time was also provided by NCSA
and the Pittsburgh Supercomputing Center.

\def\apj{Ap.\ J.}
\def\mnras{MNRAS}
\def\mn{MNRAS}


\bibliographystyle{mn2e}

\label{lastpage}
\end{document}